# *Exoplanet Science with the European Extremely Large Telescope. The case for visible and near-IR Spectroscopy at high resolution*

S. Udry[1], C. Lovis[1], F. Bouchy[2], A. Collier Cameron[3], T. Henning[4], M. Mayor[1], F. Pepe[1], N. Piskunov[5], D. Pollacco[6], D. Queloz[7,1], A. Quirrenbach[8], H. Rauer[9], R. Rebolo[10], N.C. Santos[11], I. Snellen[12], F. Zerbi[13]

[1]Astronomical Observatory of the Geneva University (Switzerland); [2]Laboratoire d'Astrophysique de Marseille (LAM, France); [3]University of St-Andrews (UK); [4]Max Planck Institute for Astrophysics (MPIA, Germany); [5]Uppsala Astronomical Observatory (Uppsala University, Sweden); [6]Warwick University (UK); [7]Cambridge University (UK); [8]Landessternwarte Heidelberg (Germany); [9]DLR (Berlin, Germany); [10]Institute of Astronomy of the Canary Islands (IAC, Spain); [11]Center for Astrophysics of the University of Porto (CAUP, Portugal); [12]Leiden Observatory (Leiden University, Netherlands); [13]Astronomical Observatory of Brera (INAF, Italy)

**Exoplanet science is booming. In 20 years our knowledge has expanded considerably, from the first discovery of a Hot Jupiter, to the detection of a large population of Neptunes and super-Earths, to the first steps toward the *characterization* of exoplanet atmospheres. Between today and 2025, the field will evolve at an even faster pace with the advent of several space-based transit search missions, ground-based spectrographs, high-contrast imaging facilities, and the James Webb Space Telescope. Especially the ESA M-class PLATO mission will be a game changer in the field. From 2024 onwards, PLATO will find transiting terrestrial planets orbiting within the habitable zones of nearby, bright stars. These objects will require the power of Extremely Large Telescopes (ELTs) to be characterized further. The technique of ground-based high-resolution spectroscopy is establishing itself as a crucial pathway to measure chemical composition, atmospheric structure and atmospheric circulation in transiting exoplanets. A high-resolution spectrograph covering the visible and near-IR domains, mounted on the European ELT, will be able to detect molecules such as water vapour, carbon dioxide and oxygen in the atmospheres of habitable planets under favourable circumstances. E-ELT HiRES is the perfect ground-based match to the PLATO space mission and represents a unique opportunity for Europe to lead the world into the era of exploration of exoplanets with habitable conditions. HiRES will also be extremely complementary to other E-ELT planned instruments specialising in different kinds of planets, such as METIS and EPICS.**

*Introduction*
The last 10 years have seen a considerable acceleration in the pace of discoveries related to exoplanets. Stunning improvements in the main detection techniques (radial-velocity measurements, transit photometry and direct imaging) have allowed the community to probe unexplored regions of parameter space. The ultra-precise HARPS spectrograph revolutionized the field by revealing the existence of a large population of super-Earths and Neptunes around nearby stars, bodies unknown to our Solar System that have masses and sizes between Earth and Neptune. NASA's Kepler mission subsequently monitored almost 200,000 stars continuously for several years at an unprecedented photometric precision. Kepler fully confirmed the existence of the low-mass planet population and obtained detailed statistical results on exoplanet orbital periods and sizes. Simultaneously, high-contrast imaging facilities obtained the first images of young giant planets on wide orbits.

One statistic may summarize the main findings from radial velocity surveys and the Kepler mission: about 50% of all stars have at least one planet with an orbital period shorter than 100 days. In particular, super-Earths and Neptunes are ubiquitous in the inner regions of planetary systems. The new frontier is now the detailed study of individual exoplanets, in particular their internal structure and atmospheric composition. Together with information on the host star and the overall planetary system architecture, there is great hope that these new insights will give us the keys to a global understanding of planet formation and evolution in our Galaxy.

*Moving from detection to characterization*
To achieve these goals, it is critical to have access to a large sample of planetary systems that are amenable to detailed characterization. Distance from the Sun plays the key role here: nearby stars are bright, their fundamental parameters including distance can be measured to high accuracy, and their exoplanets are much easier to detect and study with a variety of techniques. Nearby planetary systems have always been the focus of radial velocity and direct imaging programmes because these techniques are either photon-starved or need a sufficient angular separation between star and planet. On the contrary, transit-search experiments have focused on more distant stars because of the need to monitor a large stellar field to catch the rare transit events. The discrepancy in target samples has been identified as the major issue standing in the way of improving our knowledge of exoplanets. That is the reason why a remarkable suite of new transit-search missions and experiments are being developed:

- The Next-Generation Transit Search (NGTS) is an array of telescopes optimized for high-precision photometry that is being built at Paranal Observatory. Its main goal is to discover Neptune-mass planets and super-Earths orbiting relatively bright stars in the southern hemisphere, and will start operations in early 2015.

- The NASA K2 mission is the re-purposed Kepler satellite that now searches for exoplanet transits in various stellar fields along the ecliptic plane. The mission started earlier in 2014 and is expected to continue at least until 2016. K2 is expected to provide hundreds of new transiting exoplanets orbiting relatively bright stars.

- The NASA TESS mission (launch date 2017) aspires to detect almost all short-period super-Earths and Neptunes transiting the nearest stars to the Sun thanks to its all-sky coverage. TESS will provide excellent candidates for atmospheric characterization.

- The Swiss-ESA CHEOPS satellite (to be launched end of 2017) is a high-precision transit follow-up mission, building on transit and radial velocity surveys, whose main goal is to build a collection of precise planet radii from which accurate mean densities can be derived.

- Finally, the ESA PLATO mission will be the culmination of the transit-search efforts from space. From 2024 onwards it will thoroughly explore the solar neighbourhood and find terrestrial planets orbiting within the habitable zone of bright solar-type stars.

Besides transit search missions, several other major efforts must be mentioned:

- New high-precision radial velocity capabilities are being developed in order to provide precise exoplanet masses, particularly for the transiting planets detected by the above missions. Combining precise masses and radii yields planetary bulk densities, a fundamental observable for constraining the internal structure of exoplanets. The European spectrographs HARPS, HARPS-N and the new VLT-ESPRESSO will play a key role in this respect.

- The ESA GAIA mission (on-going) will detect a large population of giant planets orbiting a few AUs from their host star. This will significantly improve our knowledge of planetary systems at intermediate separations from the star.

- New high-contrast imaging capabilities such as VLT-SPHERE and Gemini-GPI will begin to probe the outer regions of planetary systems, particularly at young ages.

This impressive list of space missions and ground-based instruments demonstrate how strong exoplanet science has become within the broader landscape of astrophysics. While planetary systems will be explored both in their spatial extent and temporal evolution, it appears that the inner regions of nearby, mature systems will receive particular attention thanks to the power of combining the transit and radial velocity techniques. Moreover, transiting planets are amenable to a key type of observations: atmospheric characterization. Simply said, transit and occultation spectroscopy is, and will remain for a long time, the only method of studying the atmospheres of small exoplanets orbiting within 1 AU from their star, including the habitable zone.

Starting 2018, the James Webb Space Telescope (JWST) will be able to study in detail the structure and composition of the atmospheres of several Neptunes and super-Earths previously detected by the above-mentioned transit surveys. JWST will use the technique of space-based transit and occultation spectroscopy to measure exoplanet transmission and emission spectra, as pioneered successfully by the Hubble and Spitzer Space Telescopes. With respect to high-resolution and high-precision spectroscopy JWST will however have to cope with limitations intrinsic to space missions, and in particular with respect to the size of the primary mirror, the duration of the mission and thus the availability of the instruments over time, its short and long-term stability, the spectral resolution and wavelength coverage.

*The bright future of high resolution, high fidelity spectroscopy*

What is the role of Extremely Large Telescopes (ELTs) in this highly dynamic exoplanet context? Clearly, the ELTs will play a central role in atmospheric characterization, offering an extremely fruitful ground-based complementarity to JWST. ELTs have several advantages over JWST including significantly larger collecting area and higher spectral resolution. Ground-based high-resolution spectroscopy is a proven technique to obtain the transmission/emission spectrum of exoplanets during transits and secondary eclipses (Snellen et al. 2010). At high spectral resolution, absorption lines from molecular bands in the exoplanet atmosphere can be detected individually and, most importantly, planetary and telluric spectral features can be separated from each other thanks to their different apparent radial velocities. This is a critical advantage when working in spectral regions contaminated by telluric lines, and gives the ELTs a path to circumvent the difficulties of working from the ground as opposed to space for JWST. Moreover, since transits/occultations are time-critical events with a limited duration of a few hours, telescope size also plays a major role to maximize the number of photons per unit of time. The development of spectrographs on large telescopes is therefore crucial. High spectral resolution offers several other key advantages for studying exoplanet atmospheres:

- The ability to disentangle the signatures of different molecules whose transitions may overlap in wavelength, a case that could be very common. The pattern of individual molecular transitions is unique to each molecule, and even if lines cannot be detected individually, cross-correlation techniques can be used to identify and disentangle unambiguously the signatures of different molecules.
- The possibility to study physical effects that impact the intrinsic shape of atmospheric lines: wind patterns, atmospheric circulation, planetary rotation, pressure broadening, etc.
- The ability to probe high-altitude atmospheric layers in transmission, which may be undetectable at low spectral resolution because of clouds or hazes that partly suppress spectral features.

In the visible domain, transmission spectra of exoplanets may reveal the presence of a variety of atoms and molecules, among which the alkali metals Na and K, water vapour, oxygen, TiO, VO, methane and ammonia. It may also be possible to retrieve the Rayleigh scattering slope towards the blue, which is crucial to constrain the reference pressure level in the atmosphere, the nature of the main scattering species, and absolute abundances. In the near-IR, most common molecules show significant absorption bands, including water vapour, carbon monoxide, carbon dioxide, methane, ammonia, and oxygen. Broad wavelength coverage is necessary to encompass all the most interesting atmospheric signatures. On the E-ELT, the proposed METIS instrument will cover the spectral domain beyond 3 microns at high spectral resolution. The HiRES instrument discussed here will cover the range from ~0.4 to 2.4 micron and is therefore the necessary complement to METIS.

*A unique opportunity for the E-ELT and Europe*

By 2025, the ESA PLATO mission will be delivering transiting exoplanets having potentially habitable surface conditions. In the favourable case of a terrestrial planet orbiting a nearby, late-M dwarf, it has been shown that a high-resolution spectrograph on the E-ELT will be able to detect the presence of molecular oxygen in transmission using the 0.76 micron oxygen A-band (Snellen et al. 2013). Such a perspective is at the heart of the E-ELT science case. The actual detection of biosignatures on an exoplanet would have a tremendous impact on planetary science, astrophysics as a whole, and society in general. It is hardly conceivable that the E-ELT would be built without the capability of detecting such biosignatures, especially in the context of international competition. That is where the HiRES instrument enters the game in the most evident way.

Moreover, the reasons for not missing this opportunity are more general and manifold:

- HiReS remains a 'workhorse' spectrograph on a 40-m telescope. The wide range of (still un-explored) science cases has been described in the E-ELT's HiReS White Paper (Maiolino et al. 2013), including among others the variability of fundamental constants, and ultra-precise radial velocities of stars (exoplanet mass estimate).
- E-ELT HiReS is a low-risk instrument: It will work with natural seeing. It is based on proven technology and aims at demonstrated performances. No requirements are set to the light feed, e.g. the number and shape of sub-pupils. Finally it has been conceived as widely modular in terms of spectral bands and sub-instruments. This choice will allow to schedule and implement the sub-

- instruments independently and as a function of scientific priority, funding, or operational constraints.
- The in-depth characterization of exoplanets and the study of their atmospheres call for large bandwidth, high-resolution, high-fidelity spectroscopy on extremely large telescopes. E-ELT's HiReS is a unique opportunity for Europe to stay at the forefront of this emerging domain.
- Europe is going to invest considerable resources in the PLATO mission. It is crucial to consolidate this ambitious mission with adequate ground-based instrumentation. The added value arising from the combination of PLATO with HiReS can hardly be overestimated.

***Bibliography***

Maiolino, R., Haehnelt, M., Murphy, M.T., et al. 2013, arXiv:1310.3163

Snellen, I.A.G., de Kok, R.J., de Mooij, E.J.W., et al. 2010, Nature 465, 1049

Snellen, I.A.G., de Kok, R.J., le Poole, R., et al. 2013, ApJ 764, 182